\documentclass[aps,prb,preprint,amsmath,letter,showpacs,preprintnumbers,citeautoscript]{revtex4}

\usepackage{graphicx} 
\usepackage{dcolumn}
\usepackage{amssymb}
\usepackage{amsmath}
\usepackage{array}
\usepackage{longtable}

\begin{document}

\preprint{}

\title{Carrier  thermal escape  in  families of InAs/InP self-assembled quantum dots}
\author{Guillaume G\'elinas}
\author{Ali Lanacer}
\author{Richard Leonelli}
\email{richard.leonelli@umontreal.ca}
\affiliation{D\'epartement de Physique and Regroupement Qu\'eb\'ecois sur les Mat\'eriaux de Pointe (RQMP),
Universit\'e de Montr\'eal, Case Postale 6128, Succursale Centre-ville, Montr\'eal, Qu\'ebec, H3C 3J7 Canada}
\author{Remo A. Masut}
\affiliation{D\'epartement  de G\'enie Physique and Regroupement Qu\'eb\'ecois sur les mat\'eriaux de pointe
(RQMP), \' Ecole Polytechnique, Case Postale 6079, Succursale Centre-ville, Montr\'eal, Qu\'ebec, H3C 3A7 Canada}
\author{Philip J. Poole}
\author{Sylvain Raymond}
\affiliation{Institute for Microstructural Sciences, National Research Council of Canada, 1200 Montreal Road, Ottawa, Ontario, K1A 0R6 Canada}

\date{\today}
\begin{abstract}

We investigate the thermal quenching of the multimodal photoluminescence from InAs/InP (001) self-assembled quantum dots. The temperature evolution of the photoluminescence spectra of two samples is followed from 10 K to 300 K. We develop a coupled rate-equation model that includes the effect of carrier thermal escape from  a quantum dot to the wetting layer and to the InP matrix, followed by transport, recapture or non-radiative recombination. Our model reproduces the temperature dependence of the emission of each family of quantum dots  with a single set of parameters. We find that  the main escape mechanism of the carriers confined in the quantum dots is through thermal emission to the wetting layer. The activation energy for this process is found to be close to one-half the energy difference between that of a given family of quantum dots and that of the wetting layer as measured by photoluminescence excitation experiments. This indicates that electron and holes exit the InAs quantum dots as correlated pairs.

\end{abstract}

\pacs{}

\maketitle 

\section{Introduction}

Quantum dots (QDs) are designable mesoscopic atomic assemblies whose effective electronic density of states is $\delta$-functionlike.\cite{Bimberg01} One of the most studied  systems is self-assembled QDs grown in the Stransky-Krastanov mode.  It has been demonstrated that self-assembled  QDs can find applications in fields ranging from nano-optoelectronics\cite{Grundman02} to quantum computing.\cite{Kim08} Understanding the  processes that result in the thermal quenching of the photoluminescence (PL) of QDs is thus important not only on fundamental grounds but also for the realization of efficient devices operating at room temperature.

In the case of InAs/GaAs QDs, it is now well established that the two mechanisms that control the populations of electron-hole pairs in the ground state of QDs are their radiative recombination and their escape to higher lying energy states. Thermal quenching then results from  non-radiative recombination processes that occur in one or several of those higher energy states.\cite{Yang97,Sanguinetti99,LeRu03, Dawson05,Sanguinetti06,Dawson07,Torchynska07,Chahboun08,Torchynska09,Seravalli09} Even though both measurements and theoretical modeling appears straightforward, there remains to this day significant differences and apparent contradictions in the interpretation of the results published by different groups in the last decade. 

Two important questions remain unanswered. First, the identification of the higher energy states that contain non-radiative centers. Potential candidates are (i) the so-called wetting layer (WL), which is a few monolayer-thick InAs pseudomorphic quantum well  (QW) always present in the Stranski-Krastanov growth mode; (ii) the matrix  in which the QDs are inserted, either GaAs or a confining QW; (iii) defects or impurities in the matrix; or (iv) QD excited states. Second, it is still not clear whether the confined electron and holes escape a QD as a unit (exciton), as a correlated  \textit{e-h} pair or as  uncorrelated  electrons and holes. The  \textit{e-h} correlation can be evidenced by the values of the activation energies 
\begin{equation}
E^a_i=\nu \Delta E_{i} \ , \label{activation}
\end{equation} 
where $\Delta E_{i}$ is the difference between the energy of the higher energy state $i$ and that of the QDs.  If $\nu=1$, \textit{e-h} pairs escape as excitons; if $\nu=1/2$, the escape mechanism involves correlated \textit{e-h} pairs, while if $\nu<1/2$, it  involves uncorrelated electrons and holes.\cite{Yang97}

Experimentally, QD-size fluctuations result in a distribution of their quantized energies and hence to a distribution of activation energies. As temperature is increased, redistribution of the carrier population occurs towards the low energy tail of the QD energy distribution.\cite{Yang97,Sanguinetti99,Dawson05,Dawson07}  When the full-width at half-maximum (FWHM) of the emission of the ensemble of QDs is much smaller than any expected activation energy, the QD energy distribution is often represented by a $\delta$-function. The solution  of the thermal rate equations in steady-state for the ensemble photoluminescence (PL) intensity $I$ then gives
\begin{equation}
I(T) =I_0\left(1+\sum_{i=1}^m A_i\mbox{e}^{-E^a_i/kT}\right)^{-1} \ , \label{sumexp}
\end{equation}
where $m$ is the number of high energy states involved and $E^a_i$ is the activation energy for the transfer from QD to  state $i$.\cite{Torchynska07} Using this approach, Le Ru \textit{et al.} identified the GaAs matrix as the dominant recombination state and obtained $\nu=1$.\cite{LeRu03} Using samples where the QDs were imbedded in an InGaAs QW, Torchysnka \textit{et al.} found several activation energies that corressponded well to transitions from QDs to QW and from QW to the GaAs matrix, assuming $\nu=1$.\cite{Torchynska07,Torchynska09} Seravalli \textit{et al.} examined samples where the QDs were inserted in In$_x$Ga$_{1-x}$As and In$_{1-y}$Al$_y$As confining layers. They found two dominant activation energies that pointed to exciton transitions from QDs to an unidentified low-energy defect and transitions from QDs to the WL.\cite{Seravalli09}

More information is available if the internal population redistribution is also measured and modeled. In that case, the rate equations must take into account the QD energy distribution. 	Yang \textit{et al.} analysed the PL decay times  within the QD emission band with a model that only included the WL as the high energy state.\cite{Yang97} They found a good agreement with experimental data with $\nu\approx 1/2$.  Sanguinetti \textit{et al.} developed a model that included exciton transitions from QDs to WL and WL to the GaAs matrix.\cite{Sanguinetti99} It reproduced well the QD PL integrated intensity, FWHM and peak energy as a function of temperature. Dawson \textit{et al.} included in their model independant distributions of electrons and holes but they only considered transitions from QDs to the GaAs matrix.\cite{Dawson05,Dawson07} They found that the PL integrated intensity, FWHM and peak energy was best fitted by uncorrelated carrier escape.

The reason for these discrepancies is unclear. One might point out to different escape mechanisms in different samples, to insufficient information provided by a single QD emission  band, or to differences in the theoretical models and/or assumptions underlying their treatment. On the other hand, little work of a similar nature has been carried out on equivalent QDs grown on materials systems with an energy spectrum quantitatively different from that of InAs/GaAs. In order to solve some of the problems mentioned above, we present a study of the thermal quenching from two samples that contain InAs QDs embedded in an InP matrix.  Their PL spectra show several well resolved emission bands, extending from 0.75 to 1.1 eV, that can be associated with families of QDs that differ in height by one monolayer (ML).\cite{Folliot98,Poole01,Raymond03,Robertson05,Sakuma05,Michon06,Lanacer07,Dion08} The evolution with temperature of this multimodal PL  imposes stringent constraints on a model based on coupled rate equations as it should reproduce the thermal behavior of many peaks with the same set of parameters.

\section{Experimental procedures}

Sample A was grown by low-pressure metalorganic vapor phase epitaxy on (001) InP Fe-doped semi-insulating substrates at a reactor pressure of 160 torr.  After the growth at 600 $^{\circ}$C of a 100-nm InP buffer layer, the temperature was lowered  to 500 $^{\circ}$C during 90 s while still growing InP.
2.4 InAs MLs, sandwiched between two 20-ML thick InP layers, were then deposited at 500 $^{\circ}$C. The temperature was raised back to 600 $^{\circ}$C  and the growth terminated
by the deposition of a 30-nm InP cap layer. 
Trimethylindium (TMI), tertiarybutylarsine (TBA), and tertiarybutylphosphine (TBP) were used as precursors, and
Pd-purified H$_{2}$ as the carrier gas. The flow rates of TMI and TBA were kept at 0.05 and 0.95  $\mu$mol s$^{-1}$ and the TBP flow rate was 3.5 $\mu$mol s$^{-1}$.  The  growth rate of the InAs layers was close to 0.4 ML s$^{-1}$ (0.38 $\mu$m h$^{-1}$). The growth interruption sequence  used to switch from InP to
InAs and back is described in Ref.~\onlinecite{Frankland02}. In particular, a 4\,s growth interruption was applied  after InAs deposition.

Sample B was grown on (001) InP substrates by chemical beam epitaxy  from TMI, arsine (AsH$_3$), and phosphine (PH$_3$). AsH$_3$ and PH$_3$ were cracked at 850 °C in a fast switching high-temperature injector to produce predominantly As$_2$ and P$_2$. After desorption of the surface oxide, the growth was initiated with an InP buffer layer. Then, TMIn and As$_2$ were injected simultaneously into the chamber to grow about 2.2 MLs of InAs layer. This was followed by a 30 s growth interruption time under an overpressure of As$_2$ to allow the QDs to form. The QDs were then capped with InP. Further information on the growth procedure can be found in Ref.~\onlinecite{Dion08}.

The PL measurements were carried out with the samples mounted strain free in a helium-flow cryostat.
The excitation source was the 632.8 nm line of a He-Ne laser. The signal was
analyzed  with a spectrally calibrated, nitrogen-purged Fourier transform spectrometer using a liquid-nitrogen cooled InSb detector.  The PLE was excited with  the monochromatized light of a 150 W tungsten lamp. The PL was then 
 analyzed with a 0.5-m double grating spectrometer and detected with a liquid-nitrogen cooled InGaAs photodetector array sensitive up to 2 $\mu$m. The PLE spectra were not corrected by the wavelength dependence of the excitation intensity. 

\section{Results \label{results}}

\begin{figure}[h]
 \includegraphics*[width=1.0 \linewidth]{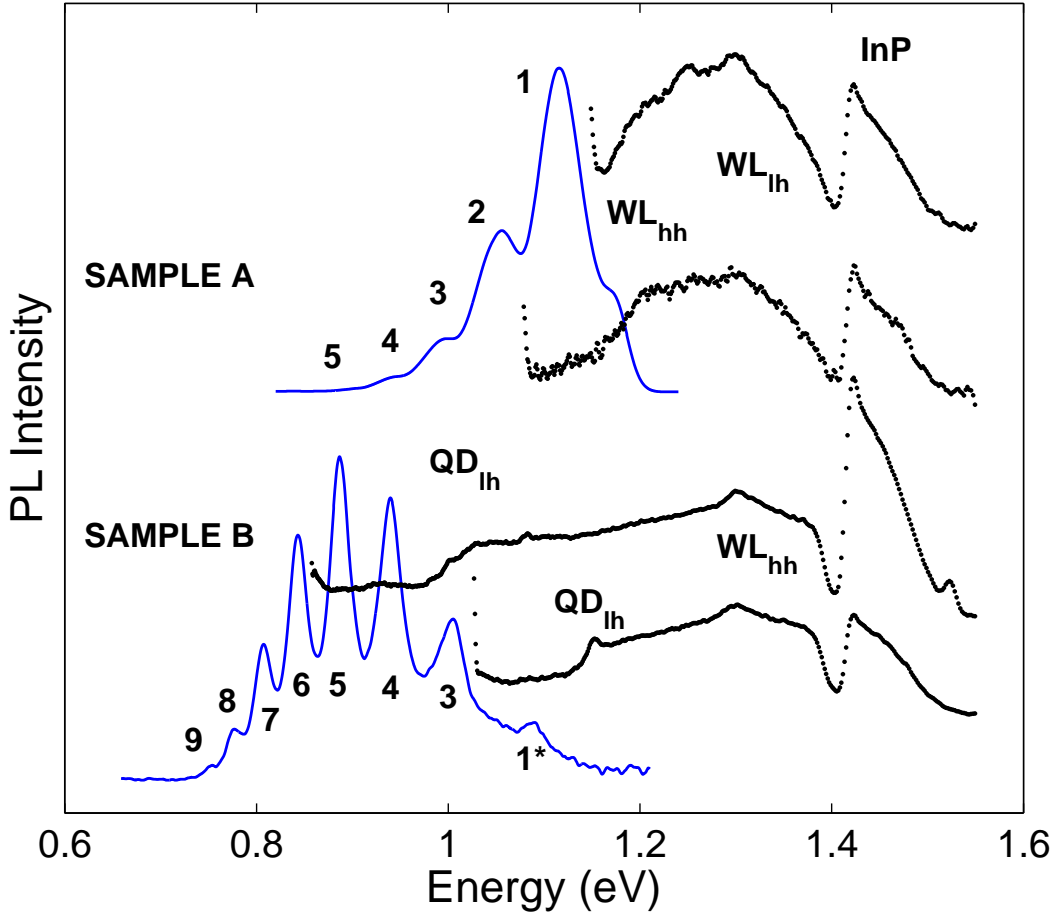}
\caption{\label{PL0} Low-temperature PL  and PLE spectra of samples A and B. QD$_{lh}$, W$_{hh}$, W$_{lh}$, and InP refer to excitonic heavy- and light-hole resonances in QDs, WL and the InP matrix respectively.}
\end{figure}

Figure \ref{PL0} shows the low-temperature PL and PLE spectra of the samples. The emission of sample A comprises five peaks while that of sample B comprises nine peaks. The energy position of peak B1$^*$ encompasses that of peaks A1 and A2 while that of peaks B3 to B5 are close to that of peaks A3 to A5. 

The PLE spectra of peak A1 shows an edge at $1.19\pm .01$ eV, labeled WL$_{hh}$,  that is close to the exciton energy in thin InAs QWs.\cite{Paki99,Lanacer07} It can thus be attributed  to the WL. It follows that the high energy tail of the emission from sample A at low-temperature  corresponds to residual emission from the WL. 

The PLE spectra of peaks B3 and B6 are also shown in fig. \ref{PL0}. The low-energy edge in both spectra can be attributed to the first excited state QD$_{lh}$. Another edge appears in both spectra at the same energy of $1.30\pm .01$ eV. Actually, this feature is common the the PLE spectra of all peaks from sample B. It can thus be associated with the WL$_{hh}$ transition in sample B. We attribute its higher energy with respect to that of the WL in sample A to the longer interruption that took place during the growth of sample B. This probably allowed the formation of thicker QDs and hence, a thinner WL with a blue-shifted resonance energy. The difference in energy  of  the fundamental WL optical transition  is useful for our purpose as it adds another constraint to the thermal model described in Section \ref{model}. 

\begin{figure}[h]
\includegraphics*[width=1.0 \linewidth]{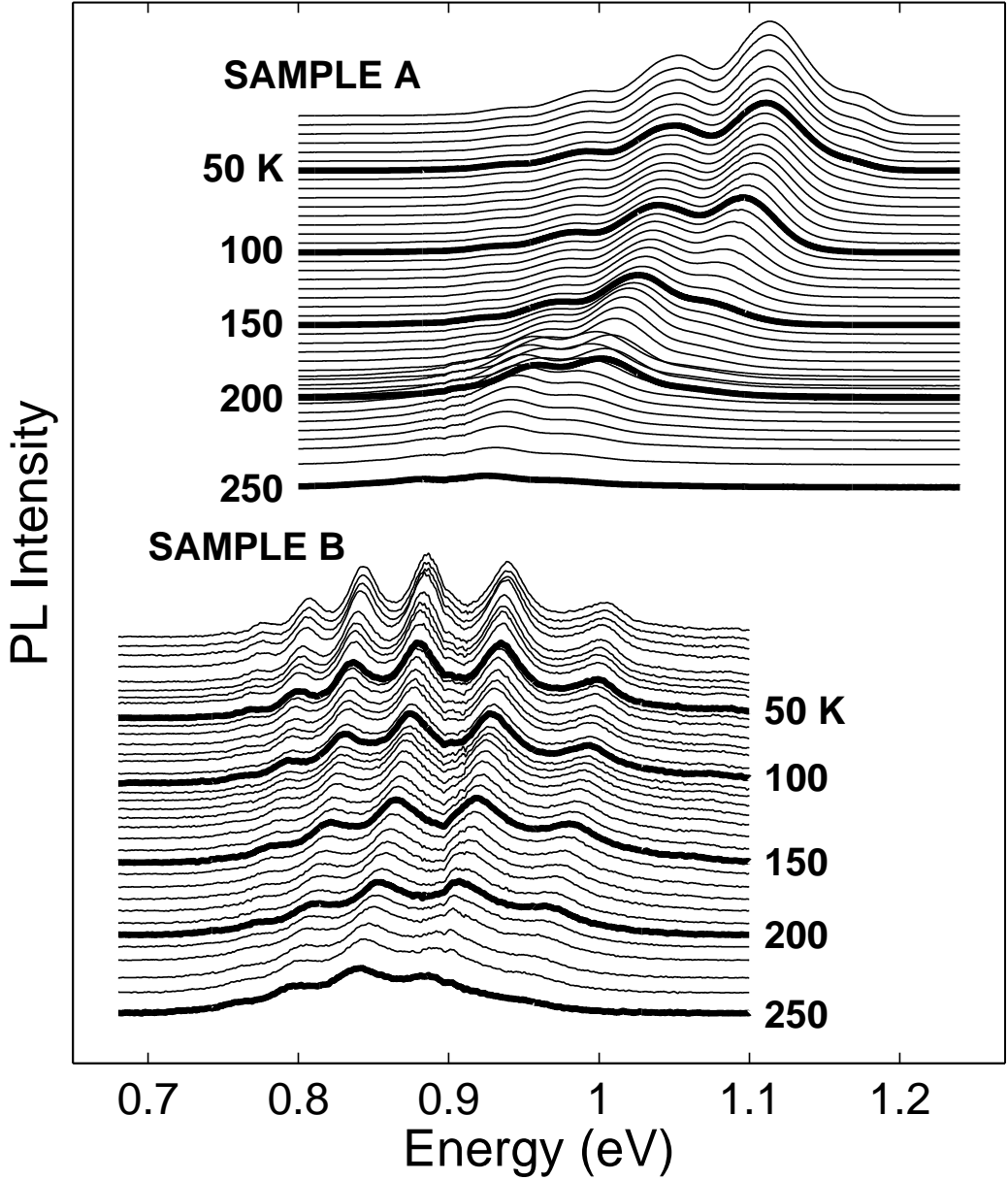}
\caption{\label{PLT} Evolution of the PL spectra of samples A and B as a function of temperature. }
\end{figure}

The evolution of the PL intensity of both samples as a function of temperature is depicted in Fig. \ref{PLT}. The emission from peak A1 is rapidly quenched for $T>100$ K while that of peak A2 remains nearly constant for $T<170$ K. The intensity of peaks A3 to A5 actually increases for $T\sim 200$ K before decreasing at higher temperatures. The emission from sample B is more robust as only peaks B1$^*$ and B3 show a significant intensity decrease at 300 K. 

\section{Rate-equation model \label{model}}

Our thermal model is  similar to those developed in Refs \, \onlinecite{Yang97} and \, \onlinecite{Sanguinetti99}. 
A series of  coupled steady-state rate equations that control the populations $n_i$ in each state $i$ is obtained from the detailed balance principle:\cite{Yang97}
\begin{subequations}
\begin{equation}
-n_i\left(R_i+\sum_{j\neq i}N_jU_{ij}\right)+N_i\sum_{j \neq i} n_j U_{ji} +P_i =0\, ,
\end{equation}
where $R_i$ is the recombination rate of state $i$, $N_i$  the number of states per unit area, and $U_{ij}$  the transfer cross-section from state $i$ to state $j$. $P_i$ represents the carrier generation, $\nu_i$ is defined in Eq. (\ref{activation}) and
\begin{equation}
U_{ji} = U_{ij}\exp\left\{\nu_i\dfrac{(E_i-E_j)}{kT}\right\} \quad \text{if } E_i>E_j \, .
\end{equation}
\end{subequations}

We improved on previous models by incorporating several unique features.

\noindent \textit{(i)} As both WL and InP matrix features are observed in the PLE spectra,  transfer from QDs to WL, from QDs to the InP matrix, and from WL to the InP matrix are allowed. 

\noindent \textit{(ii)} The allowed states of  the WL and the InP matrix are distributed over a wide range of energy above their fundamental edge, $E_g^W$ for the WL and $E_g^M$ for InP. To make this fact numerically tractable, the WL is separated into segments of width $\Delta E_{W}$ extending from $E_g^{W}$ to $E_g^{M}$. The effective number of states per unit area of each segment is $N_{W}=D_{W} \Delta E_{W}$, where $D_{W}=m_x^{InAs}/\pi\hbar^2$ is the two-dimensional density of states in the WL. The InP matrix is similarly segmented. We assume a $\sqrt{E-E_g}$ dependence for its three-dimensional density of states. The number of states per unit area for a segment that extends from $E_i^{min}$ to $E_i^{min}+\Delta E_{M}$  is thus 
\begin{equation}
\begin{split}
N_{Mi}=\frac{2}{3}D_M\ell_M [&(E_i^{min}+\Delta E_{M}-E_g^{M})^{3/2}\\
&-(E_i^{min}-E_g^{M})^{3/2}] \, , \label{densite}
\end{split}
\end{equation}
 where $D_M=2^{1/2}(m_x^{InP})^{3/2}/(\pi^2\hbar^3)$ (Ref. \, \onlinecite{Yu04}) and $\ell_M$ is the active thickness of the matrix.
  
\noindent \textit{(iii)} $E_g^{W}(T)$ and $E_g^{M}(T)$ are assumed to follow the Varshni temperature dependence   with the parameters of bulk InAs and InP, respectively.\cite{Madelung04} 

In order to restrain the number of adjustable parameters, the following assumptions were made.
\begin{subequations}\label{parameters}
\begin{eqnarray}
R_i&=&\left\{
\begin{array}{l l}
R_D  &  \mbox{if $ i \in$ $D$,} \\
R_W &   \mbox{for  $W$ lowest energy segment,}\\
R_M &   \mbox{for  $M$ lowest energy segment,}\\
0 &   \mbox{for all other states,}\\
\end{array} \right. \label{R}\\
\nu_{i}&=&\left\{
\begin{array}{l l}
\nu_D &  \mbox{if $i \in$ $D$,}\\
1 &  \mbox{if $i \in$ $W$,}\\
\end{array} \right. \\
P_i&=&\left\{
\begin{array}{l l}
P &  \mbox{if $i= M$ highest energy segment,}\\
0 &  \mbox{if not,}\\
\end{array} \right. \\
U_{ij}&=&\left\{
\begin{array}{l l}
0 & \quad \mbox{if $i$ and $j\in$ $D$,}\\
U_{WD} &  \mbox{if $i\in$ $W$ and $j\in$ $D$,}\\
U_{MD}&  \mbox{if $i\in$ $M$ and $j\in$ $D$,}\\
U_{MW}&   \mbox{if $i\in$ $M$ and $j\in$ $W$,}\\
U_{WW}&  \mbox{if $i$ and $j\in$ $W$,}\\
U_{MM}&  \mbox{if $i$ and $j\in$ $M$,}\\
\end{array} \right. \label{U}
\end{eqnarray}
\end{subequations}
where $D$, $W$, and $M$ refer respectively to the ensemble of QD, WL, and InP matrix states.
We further assumed that the parameters in Eqs \ref{parameters} are independant of temperature. 

\begin{figure}[h]
\includegraphics*[width=0.7 \linewidth]{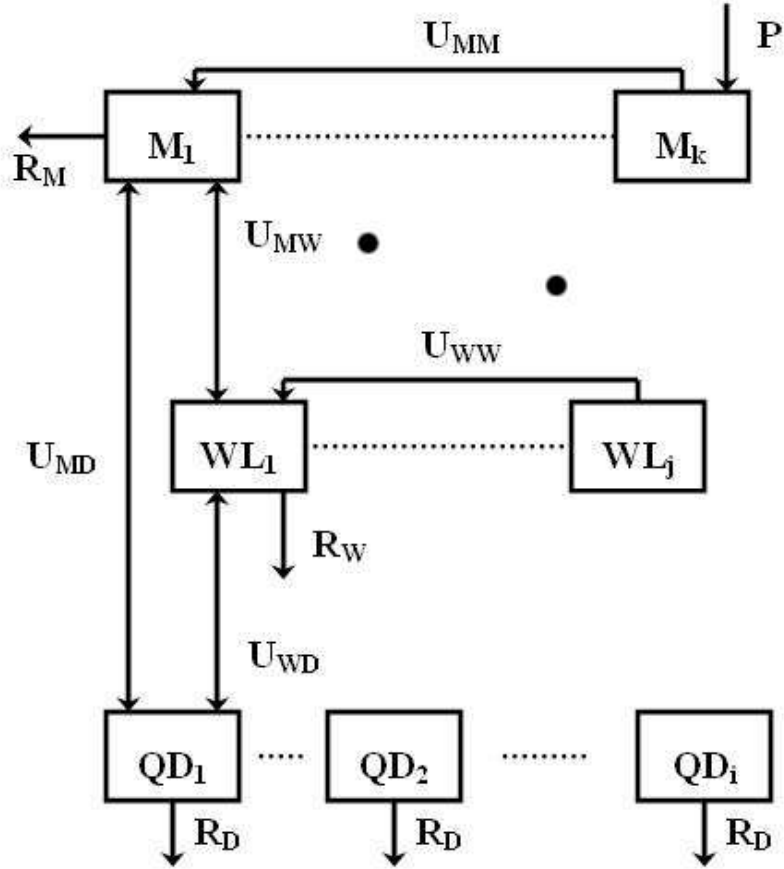}
\caption{\label{scheme} Schematic of the coupled rate-equation model.}
\end{figure}

In Eq. \ref{R}, $R_D$ corresponds to QD radiative rate. As no emission from the WL nor the InP matrix is observed at high temperature, $R_W$ and $R_M$ correspond to non-radiative rates. $R_W$ was  assigned only to the lowest energy segment because, in a QW, free excitons form 2-dimensional polaritons that  do not couple to photon-like polaritons propagating perpendicular to the QW plane unless their energy is within a small bandwidth near the bottom of the band.\cite{Damen90} Excitons must thus relax to the bottom of their energy band before they can recombine radiatively or non-radiatively.\cite{Leonelli93} A similar argument can be made for $R_M$.\cite{Steiner86} Figure \ref{scheme} schematizes the rate-equation model used to analyze our data.

\section{Discussion}

\begin{table*}
\caption{\label{optim} Parameters obtained by fitting the model to the experimental data. The numbers in parenthesis are the uncertainties calculated with Eq. \ref{erreur}.}
\begin{ruledtabular}
\begin{tabular}{lcccccccc}
& $R_W/R_D$ & $R_M/R_D$ & $\ell_M$ & $U_{WD}D_0/R_D$ & $U_{MD}D_0/R_D$ & $U_{MW}$ & $\nu_D$ & $E_g^{W}(0)$ \\
&  & &(nm) & & & (Hz cm$^2$)& & (eV)\\
\hline
Sample A &22 (1)&9 (18)& 100\footnotemark[1]&12.3 (0.5) &5000 (300)&0.71 (0.03)&0.64 (0.03)&1.19 (0.04)\\
Sample B & 36 (2)&20 (10)&100\footnotemark[1]&51 (2)&20 (140)&10 (10)&0.51 (1) &1.30\footnotemark[1]
\end{tabular}
\end{ruledtabular}
\footnotetext[1]{Fixed parameter.}
\end{table*}

To compare the simulations with the experiments, the data were treated as follows. The peak energy and integrated intensity of each of each peak at a given temperature was obtained by fitting the PL spectrum  with a series of gaussians peaks. This procedure was found to reproduce well the PL spectra except for peak B1$^*$. Its peak energy and intensity was obtained by substracting the intensity of all the other peaks from the total intensity of the PL emission. The energy position of each family of QDs served as input to the model. 

\begin{figure}[t]
 \includegraphics*[width=1.0 \linewidth]{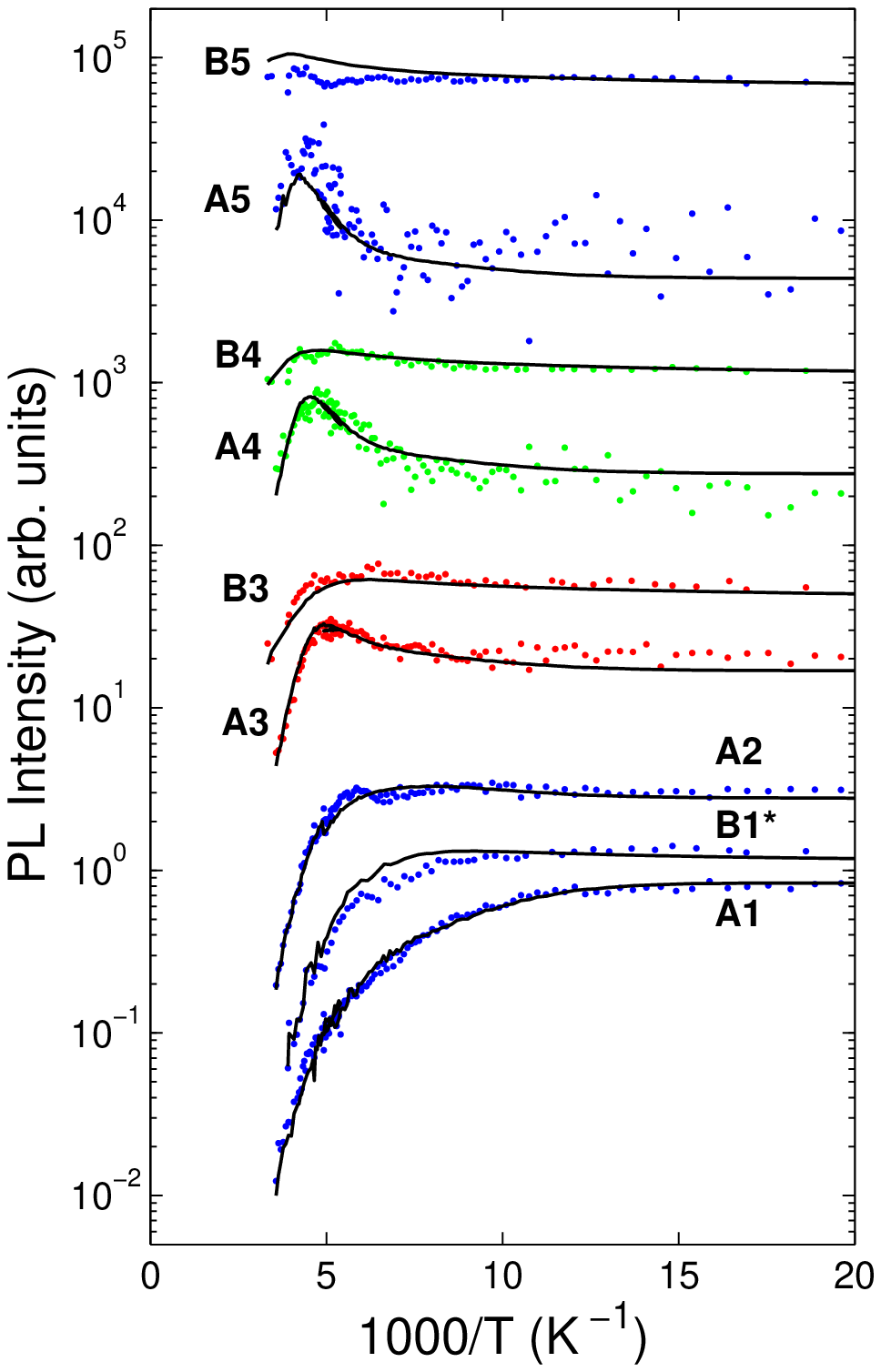}
\caption{\label{arrh} Arrhenius plot of the temperature dependence of the integrated intensity of the individual families of QD's . Labels A and B identify peaks from samples A and B respectively. The curves have been shifted vertically for clarity.}
\end{figure}

In the model, the total number of QD states per unit area $D_0$ and the QD recombination rate $R_D$ are scaling factors. The relative number of states for each QD family was assumed to be given by the relative intensity of the PL at low temperature. $\Delta E_W$ and $\Delta E_M$ were set at 10 meV, a value close to the spectral extent of the absorption edge.   In Eq. \ref{U}, the parameters $U_{WW}$ and $U_{MM}$ were set to a high value to ensure that the excitons in the WL and the InP matrix are in thermal equilibrium. Finally, we fixed $\ell_M$ at 100 nm, a value close to the penetration length of the excitation source.\cite{note1} The model is thus left with seven  adjustable materials parameters:  $R_W/R_D$, $R_M/R_D$,  $U_{WD}D_0/R_D$, $U_{MD}D_0/R_D$, $U_{MW}$, $\nu_D$, and $E_g^{W}(0)$.

The result of our simulations is presented in Fig. \ref{arrh}. The temperature dependence of all peaks from sample A are very well reproduced by our model with the optimized parameters listed in Table \ref{optim}. In particular, our model reproduces the intensity increase of peaks A3 to A5 when $T\agt 180$ K. The uncertainties $\Delta a_i$ of the optimized parameters were estimated using\cite{Bevington03} 
\begin{equation}
\Delta a_i=\left(\dfrac{\partial \chi^2}{\partial a_i}\right)/\left(\dfrac{\partial^2 \chi^2}{\partial^2 a_i}\right) \, . \label{erreur}
\end{equation}
Stricktly speaking, Eq. \ref{erreur} is only valid if the cross-partial derivatives $\partial \chi^2/\partial a_i \partial _j$ are small with respect to the diagonal terms, which is not the case here. However, it gives a good estimate of the sensitivity of the fit to a given parameter.

  It can be seen from Table \ref{optim} that the parameters with high uncertainties are  relative to the InP matrix, an indication that the main QD escape channel is through the wetting layer.  To further test this hypothesis, we deactivated the contribution to the thermal quenching  of InP matrix by  fixing $R_M=0$, with no significant change to the fit. On the other hand, no fit could be achieved when the the  WL was similarly deactivated. Further, the fitted value for $E_g^{W}(0)$ corresponds within uncertainties to the measured value of the low energy edge $WL_{hh}$ shown in Fig. \ref{PL0}. We can thus conclude that in sample A, the main quenching mechanism is through carrier escape from QD to WL  followed by a non-radiative recombination of the carriers in the WL.
  
There is globally much less thermal quenching in sample B and thus less dynamics to constrain the model. To extract relevant information, we fixed $E_g^{W}(0)$ to the value of $WL_{hh}$ obtained  the PLE spectra of sample B, as shown in Fig. \ref{PL0}. Here also, peaks B1$^*$, B3, and B4 are well reproduced by our simulations. There was nearly no change of intensity for peaks B5 to B9, while our model predicts a slight increase. The discrepancy can easily be explained by our neglect of the temperature dependence of the parameters of the model. We also note that the two most significant materials parameters, $R_W$ and $U_{WD}$ have realistic and comparable values for both samples. 

As for the parameter $\nu_D$, the simulations indicate that it is close to one-half for our samples. Further, our model could not reproduce the data  if $\nu_D$ was fixed to 1. This  indicates that electron and holes escape from the QDs mostly as correlated \textit{e-h} pairs. We thus corroborate the findings of Yang \textit{et al.}\cite{Yang97}

It is instructive to simulate with our model the temperature dependence of the integrated intensity of a monomodal QD emission. We have generated a gaussian distribution of fifteen QD subfamilies centered at an energy $E^D$ and shifted with respect to the WL by  $\Delta E=E_g^W-E^D$. The FWHM of the distribution was fixed at $0.25 \Delta E$, a value typical of monomodal InAs/GaAS QD emission.\cite{LeRu03} We used in the simulations the same materials parameters as those found for sample A except for $\nu_D=0.5$. 
 
\begin{figure}[h]
 \includegraphics*[width=0.80 \linewidth]{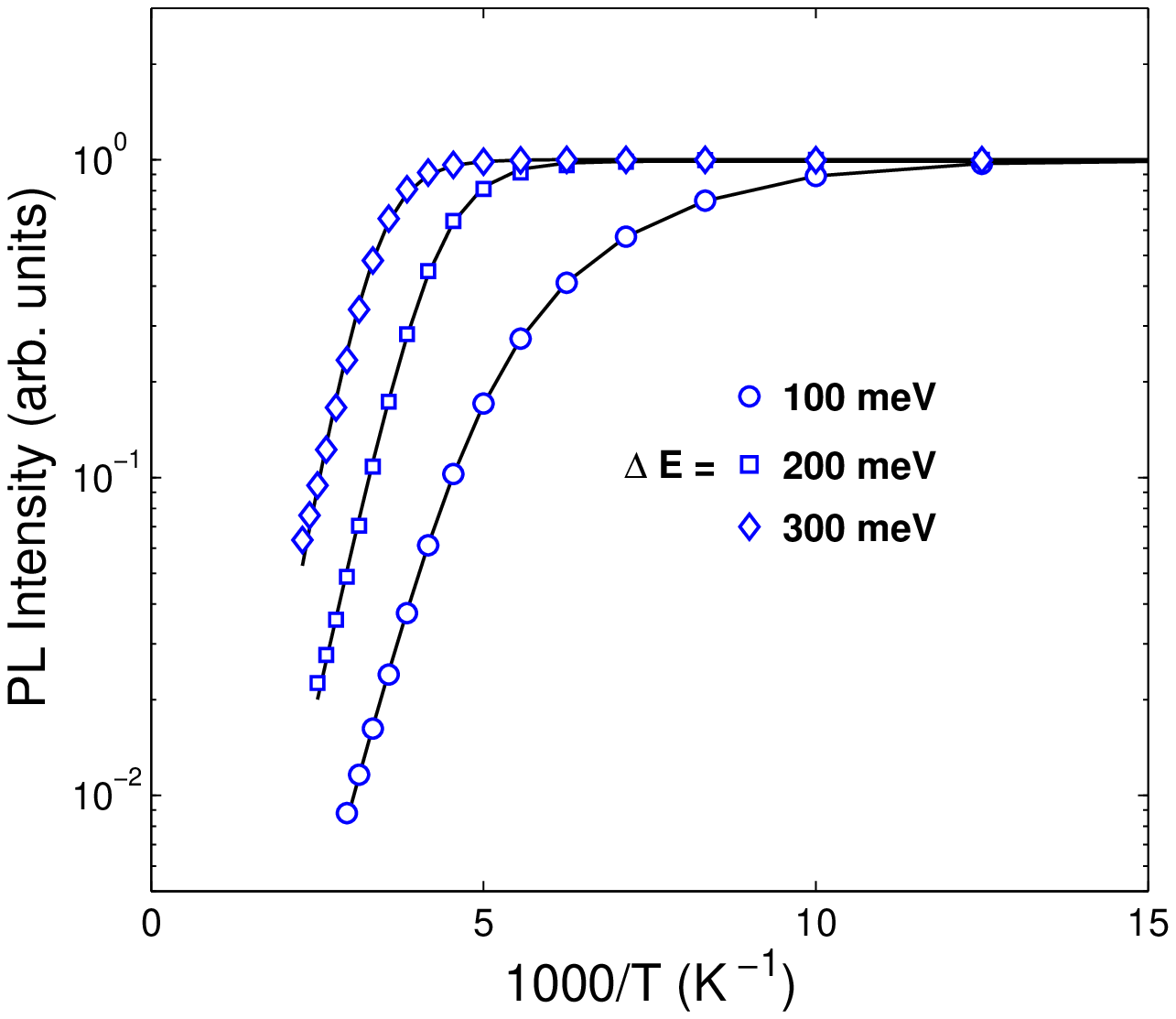}
\caption{\label{arrhtest} Arrhenius plot of the temperature dependence of the integrated intensity of simulated monomodal QD emission for three values of $\Delta E=E_g^W-E^D$. The solid lines are best fits using Eq. \ref{sumexp}}
\end{figure}

The result of our simulations for $\Delta E=$ 100, 200, and 300 meV is shown as symbols in Fig. \ref{arrhtest}. The curves were analyzed with a sum of activated processes as described by Eq. \ref{sumexp}. All curves are well reproduced with the activation energies  given in Table \ref{testtable}.   The rapid drop of intensity at elevated temperatures is controlled by $E_1^a$. In all three cases, we find $E_1^a>\nu\Delta E$. Further, $E_1^a$ does not correspond to any difference between the energy levels present in the model. 

\begin{table}[h]
\caption{\label{testtable} Parameters obtained by fitting the simulated data of Fig. \ref{arrhtest} with Eq. 
\ref{sumexp}.}
\begin{ruledtabular}
\begin{tabular}{ccc}
$\Delta E$ & $E_1^a$& $E_2^a$\\
(meV)&(meV)  &(meV)\\ 
\hline
100 & 160 & 50\\
200\footnotemark[1] & 190 &  \\
300\footnotemark[1] & 230  \\
\end{tabular}
\end{ruledtabular}
\footnotetext[1]{Only one activation energy required.}
\end{table}

These simulations show that the difference between $E_1^a$ and the activation energies inserted in the model comes from  carrier transport in the WL and recapture by QDs. This induces a redistribution of the carriers within the subfamilies that slows down thermal quenching, resulting in an ensemble effective activation energy  higher than actual ones. Therefore, in systems where recapture competes with recombination,  Eq. \ref{sumexp} gives empirical activation energies that might not correspond to any physical process at play.

\section{Conclusions}

We have developed a system of coupled rate equations for the temperature dependence of the  multimodal PL of InAs/InP QDs. The model includes carrier escape to the InAs wetting layer and to the surrounding InP matrix as well as carrier transport and retrapping. Even though our model comprises seven adjustable parameters, the constraints imposed by the simulation of the complex temperature behavior of up to five different QD families makes our fits robust. We find that the main quenching mechanism is induced by carrier escape to the wetting layer followed by non-radiative recombination.  Further, our results clearly establish that,  for both samples examined,  electrons and holes are emitted as correlated pairs rather than excitons. Finally, we show that carrier redistribution within the QD energy levels as temperature in increased can yield activation energies obtained from analyzing  PL integrated intensities that do not correspond to any actual physical process.

We cannot assert whether correlated-pair escape is characteristic of self-assembled QDs or specific to our samples. The latter case could mean that the temperature dependence of QD optical emission is governed by microscopic parameters such as the size and shape of individual QDs. A better theoretical understanding of the interactions between QDs and their environment is thus not only of great fundamental interest, but could also impact the design of QD-based devices. 
 
\begin{acknowledgments}

This work was supported by the Natural Sciences and Engineering Research Council of Canada (NSERC) and the Fonds
Qu\'eb\'ecois de la Recherche sur la Nature et les Technologies (FQRNT). 

\end{acknowledgments}


\begin{thebibliography}{30}
\bibitem[{not()}]{note1}
\bibinfo{note}{Simulations were also run with $\ell_M$ as an adjustable
  parameter without any significant difference.}
\end{thebibliography}

\end{document}